\documentclass[usenatbib]{mn2e}

\usepackage{graphicx}
\usepackage{amssymb}
\usepackage{epsfig}

\title[INTEGRAL-SMC]{INTEGRAL deep observations of the Small Magellanic Cloud}

\author[M.J. Coe et al.]{
M.J. Coe$^{1}$,
A. J. Bird$^{1}$,
D.A.H. Buckley$^{2}$,
R.H.D. ~Corbet$^{3}$,
A.J. Dean$^{1}$,
\and
M. Finger$^{4}$,
J.L. Galache$^{5}$,
F. Haberl$^{6}$,
V.A. ~McBride$^{1}$,
I. Negueruela$^{7}$,
\and
M. Schurch$^{1,8}$,
L.J. Townsend$^{1}$,
A. Udalski$^{9}$,
J. Wilms$^{10}$,
A. Zezas$^{11}$ \\
$^{1}$ School of Physics and Astronomy, University of Southampton, SO17 1BJ, UK\\
$^{2}$ South African Astronomical Observatory, PO Box 9, Observatory 7935, Cape Town, South Africa\\
$^{3}$ University of Maryland, Baltimore County, Mail Code 662, NASA Goddard Space Flight Center, Greenbelt, MD 20771, USA \\
$^{4}$ National Space Science and Technology Center, 320 Sparkman Drive, Huntsville, AL, USA \\
$^{5}$ Harvard-Smithsonian Center for Astrophysics, 60 Garden St, Cambridge, MA 02138, USA \\
$^{6}$ Max-Planck-Institut fur extraterrestrische Physik, Giessenbachstrasse, 85748 Garching, Germany \\
$^{7}$ Departamento de Fisica, Ingenieira de Sistemas y Teoria de la Seal, Universidad de Alicante, 03080 Alicante, Spain\\
$^{8}$ Astrophysics, Cosmology and Gravity Centre (ACGC), University of Cape Town, Private Bag, X3 Rondebosch 7701, South Africa\\
$^{9}$ Warsaw University Observatory, Al. Ujazdowskie 4, 00-478 Warszawa, Poland \\
$^{10}$ Dr. Karl Remeis-Observatory \& ECAP, University of Erlangen-Nuremberg, Sternwartstr. 7, 96049 Bamberg, Germany \\
$^{11}$ IESL, Foundation for Research and Technology, 71110 Heraklion, Crete, Greece \\
}

\begin{document}

\date{12 Apr 2010}

\pagerange{\pageref{firstpage}--\pageref{lastpage}} \pubyear{2002}

\maketitle

\label{firstpage}

\begin{abstract}

Deep observations of the Small Magellanic Cloud (SMC) and region were carried out in the hard X-ray band by the INTEGRAL observatory in 2008-2009. The field of view of the instrument permitted simultaneous coverage of the entire SMC and the eastern end of the Magellanic Bridge. In total, INTEGRAL detected seven sources in the SMC and five in the Magellanic Bridge; the majority of the sources were previously unknown systems. Several of the new sources were detected undergoing bright X-ray outbursts and all the sources exhibited transient behaviour except the supergiant system SMC X-1. They are all thought to be High Mass X-ray Binary (HMXB) systems in which the compact object is a neutron star.

\end{abstract}


\section{Introduction and background}

The Small Magellanic Cloud (SMC) present a nearby, easily observable
irregular galaxy in which it is possible to study the results of tidal interactions of star formation. Multiwavelength studies of the SMC have shown that
it contains a large number of X-ray binary pulsars  all but
one of them in Be/X-ray binary systems (Coe et al. 2005,
2009). This preponderance of such systems is most likely
explained by tidally-triggered star formation induced
by the most recent close approach of the SMC and LMC
(Gardiner \& Noguchi 1996). However, the most recent close
approach of the SMC and LMC was ~200 Myr ago  - much
longer than the evolutionary timescale of Be/X-ray binaries.
This implies that either there has been a significant delay
between the encounter of the SMC and LMC and the onset
of star formation, or that subsequent waves of star formation
have given rise to these Be/X-ray binaries (Harris \&
Zaritsky 2004). These objects, being tracers of star formation
(Grimm et al. 2003), give direct insights into the star
formation history of their host galaxies.

In this paper an extended set of deep observations of the SMC region are reported.
One major new discovery from this INTEGRAL study was a population of
HMXBs in the Magellanic Bridge. These observations have been published
elsewhere(McBride et al, 2010). This paper focusses exclusively on the HMXB population of the SMC itself.

\section{Observations}

The IBIS telescope (Ubertini et al. 2003) on board INTEGRAL,
which is optimised for an energy range of 15--200 keV
and has a field of view of 29$\times$29 degrees (Full Width Zero Intensity response), is well suited to
observing large sky areas for point sources. As part of a
key programme monitoring campaign on the SMC and 47
Tuc, INTEGRAL observed the SMC and Magellanic Bridge
for approximately 90 ks per satellite revolution ($\sim$3 days)
from 2008 November 11 to 2009 June 25 - see Table~\ref{tab:expo} for a detailed journal of all the observation dates and exposures.

\begin{table}
\caption{INTEGRAL observation log.}
\label{tab:expo}
 \begin{tabular}{llll}
  \hline
   Rev Number & MJD$_{\rm start}$ & Exposure (ks) &  \\
  \hline
745 & 54788.7 & 43.8 & \\
 746 & 54791.8 & 47.6 & \\
 747 & 54794.8 & 47.2 & \\
 748 & 54797.7 & 46.8 & \\
 749 & 54800.8 & 48.0 & \\
 750 & 54803.7 & 55.7 & \\
 751 & 54806.7 & 64.6 & \\
 752 & 54809.7 & 57.7 & \\
 753 & 54812.7 & 58.8 & \\
 754 & 54815.7 & 45.3 & \\
 755 & 54818.7 & 60.9 & \\
 756 & 54821.7 & 57.2& \\
 796 & 54941.4 & 79.6 & \\
 797 & 54944.4 & 62.7 & \\
 812 & 54989.2 & 63.3 & \\
 813 & 54992.2 & 53.1 & \\
 814 & 54995.2 & 60.8 & \\
 815 & 54998.2 & 58.8 & \\
 816 & 55001.2 & 62.1 & \\
 817 & 55004.2 & 63.1 & \\
 818 & 55007.2 & 40.2 & \\
 \hline
 \end{tabular}
\end{table}

Individual pointings (science windows) were processed
using the INTEGRAL Offline Science Analysis v.7.0 (OSA,
Goldwurm et al. 2003) and were mosaiced using the
weighted mean of the flux in the 3--10 keV (JEM-X) and
15--35 keV (IBIS) energy ranges. Proprietary software was
used to mosaic the observations from successive revolutions
to improve the exposure and thereby the sensitivity to faint
sources. Lightcurves in these energy bands were generated
on science window (~2000 s) and revolution time-scales. The
IBIS energy band was chosen to maximise the detection significance
of SMC X-1, and hence other SMC accreting X-ray
pulsars, which have similar spectral shapes to SMC X-1 in
this energy range. An IBIS mosaic of data from revolutions
752--756 in the 15--35 keV is shown in McBride et al (2010).

\section{Individual Sources seen in the SMC}

A list of all the definite source detections and when they occurred is presented in Table~\ref{tab:sources}.

\begin{table}
  \caption{Table of definite INTEGRAL source detections in the SMC}
  \label{tab:sources}
\begin{tabular}{|c|c|}
  \hline
  Source name & Times of detection (MJD) \\
  \hline
  SMC X-1 & Various - see  Figure~\ref{Fig:smcx1_revlc}\\
  SXP11.5 & 54989 - 55007 \\
  SXP756 & 54941 - 54947 \\
  IGR J00515-7328 & 54800 - 55000\\
  SXP6.85 & 54790 - 54820 \\

  \hline

\end{tabular}
\end{table}

\subsection{SMC X-1}

As a neutron star accreting from a supergiant overfilling its Roche lobe, SMC~X-1 is the only persistent accreting X-ray pulsar in the SMC.  This transfer of angular momentum serves to spin up the pulsar, and SMC X-1 has demonstrated consistent spin up over the last 40 years.  It has a 3.89\,d orbital period with eclipses lasting $1.12$\,d (Wojdowski et al. 1998) and also shows superorbital modulation at $\sim60$\,d, thought to be the result of a precessing, warped accretion disk (Clarkson et al. 2003).

\begin{figure}
\centering
\includegraphics[width=90mm]{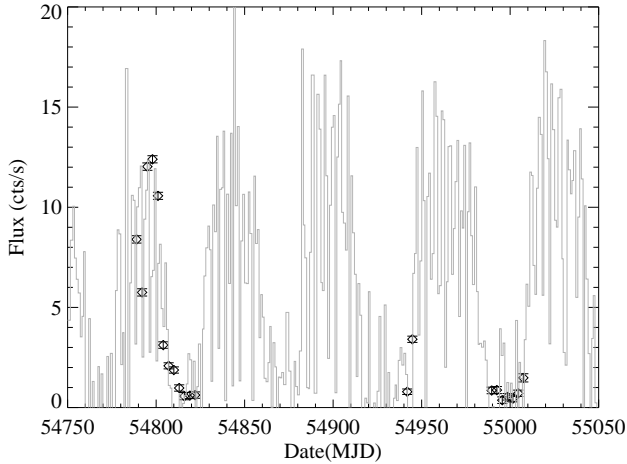}
\caption{The IBIS measurements of SMC X-1 (diamonds). Each measurement represents the flux (15--35\,keV) averaged over a spacecraft orbit.  The RXTE/ASM lightcurve, arbitrarily scaled is overplotted in grey to show the superorbital modulation and the binary eclipses. }
\label{Fig:smcx1_revlc}
\end{figure}

\begin{figure}
\centering
\includegraphics[width=80mm]{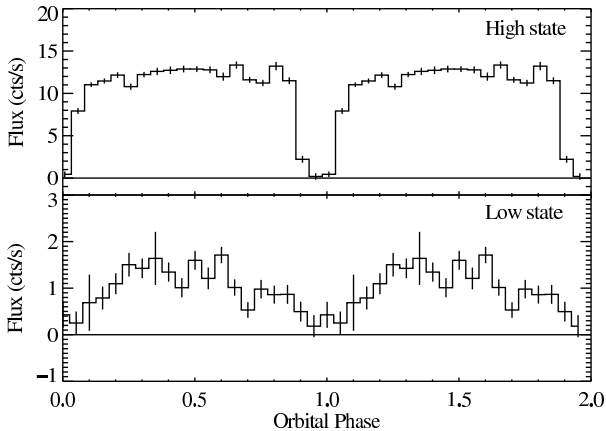}
\caption{.  IBIS 15-35 keV data of SMC X-1 folded on the orbital period during superorbital high state (top panel) and superorbital low state (lower panel).  The binning is 0.05 of the orbital period.}
\label{Fig:smcx1_ecl}
\end{figure}

The first set of IBIS observations cover almost a full superorbital cycle, while the subsequent observations occur predominantly in the superorbital low state.  See Fig.~\ref{Fig:smcx1_revlc} for the whole lightcurve and Figure~\ref{Fig:smcx1_ecl} for details of the binary eclipse.

Folding data from revolutions 745 to 749 on the orbital ephemeris from Wojdowski et al. (1998) resulted in an orbital profile during superorbital high state.  (See Figure~\ref{Fig:smcx1_ecl} top panel).  For comparison, an orbital profile during superorbital low state was generated using data from revolutions 751 to 756 and 812 to 818.  There are significant differences between the eclipse profiles during these two superorbital phases.  Eclipse ingress and egress are steep, almost square, during superorbital high state, while they are much more gradual (almost 0.1 phase longer) during superorbital low state.  These results are in agreement with the orbital profile as a function of superorbital phase presented in Trowbridge, Nowak \& Wilms (2007).  These authors use softer X-ray data (0.2-12\,keV) to show that the eclipse is wide and shallow during superorbital low state while it is narrow and deep during superorbital high state.

From these data there is no evidence that the eclipse in hard X-rays (15-35\,keV) is shorter than in soft X-rays (0.2-12\,keV, Trowbridge, Nowak \& Wilms 2007), which may be expected due to the increased transparency of the supergiant photosphere to hard X-ray photons.  There may, however, be some evidence for asymmetry of ingress and egress during the high superorbital state eclipse.  The eclipse profile in Figure~\ref{Fig:smcx1_ecl} shows a similar shape to 20-40\,keV eclipse profile in McBride et al. (2007).

\begin{figure}
\includegraphics[width=87mm]{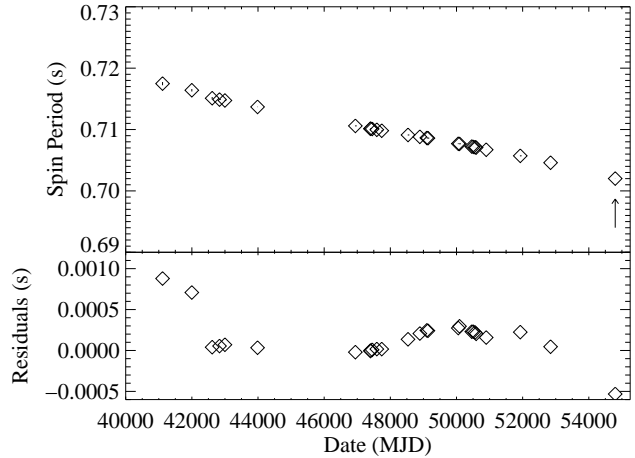}
\caption{Upper panel: The history of the spin period of SMC X-1. The latest IBIS measurement on MJD 54797 is indicated by an arrow.The penultimate measurement is from the previous INTEGRAL observations in 2003 (McBride et al, 2007) Lower panel: the deviations of the period from a simple linear trend.}
\label{fig:smc2}
\end{figure}

An IBIS observation at MJD 54797, during the superorbital bright phase, was used to determine the current spin period of the pulsar in SMC X-1.  This was determined using an epoch folding technique to be $0.702010\pm0.000008 $\,s, and is plotted above the arrow in Fig~\ref{fig:smc2}.  A straight line model does not fit the period evolution very well, but gives a good idea of the general spin up of the system over its lifetime, which can be estimated as $\sim-3.8\times10^{-4}$\,s/yr.

\subsection{SXP756}

\begin{figure}
\includegraphics[width=70mm,angle=90]{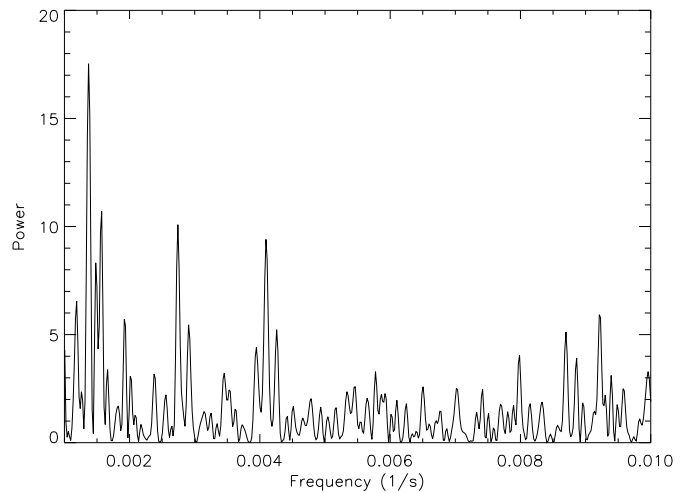}
\caption{Lomb-Scargle periodogram of RXTE PCA data for SXP756 covering the period range 100-1000s. The highest peak corresponds to a period of 730$\pm$5s.}
\label{fig:756ls}
\end{figure}

A source was detected in INTEGRAL at a position coincident with SXP756 (Coe \& Edge, 2004) within an uncertainty of $\sim$1 arcminute over two satellite orbits (MJD54941 - 54947). Follow-up RXTE observations (MJD54943)
revealed evidence for weak pulsations around 730$\pm$5s, the relatively large uncertainty reflecting the fact that the observation extended over 13ks and therefore does not cover many period cycles.
Figure~\ref{fig:756ls} shows the Lomb-Scargle periodogram for the RXTE/PCA data.

\begin{figure}
\includegraphics[width=80mm,angle=-0]{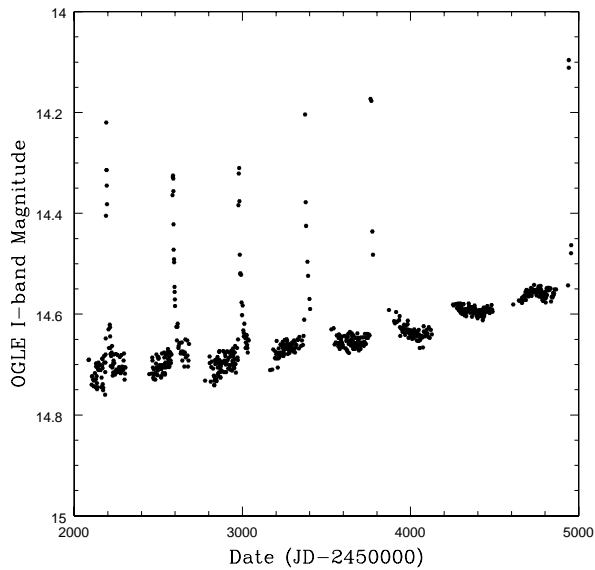}
\caption{OGLE III I-band lightcurve of the optical counterpart to SXP756. Optical flares are clearly visible every 394d (Coe \& Edge, 2004). The INTEGRAL detection was coincident with the last optical flare seen in these data.}
\label{fig:ogle1}
\end{figure}

Independent confirmation that both the INTEGRAL and the RXTE sources are probably SXP756 comes from the OGLE III optical monitoring
of the counterpart. This source is unusual in that it shows striking optical behaviour around the time of the
periastron passage of the neutron star (Coe \& Edge, 2004). These authors quote an ephemeris for outbursts
of T = JD2449830 + 394N. The outburst reported here peaked at JD2454945 which gives a phase for the INTEGRAL observations of 0.98 using the published ephemeris; coinciding with the sharp optical flare seen in the OGLE III data at that time - see Figure~\ref{fig:ogle1}. Thus there can be no doubt that it is SXP756 that was seen in the INTEGRAL data.

\begin{figure}
\includegraphics[width=80mm,angle=-0]{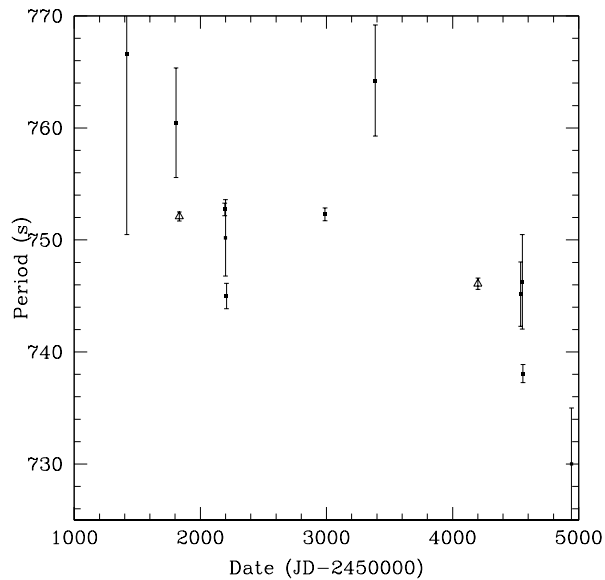}
\caption{The pulse period history of SXP756. All the measurements shown, except the two marked with a triangle, are from RXTE. The triangle data points are from XMM (Haberl \& Pietsch 2004, Haberl, Eger \& Pietsch 2008).The final point coincides with the INTEGRAL detection reported in this work.}
\label{fig:756period}
\end{figure}

The significant neutron star spin up in this system is shown from our long-term RXTE monitoring programme - see Figure~\ref{fig:756period}. Measurements up to MJD54000 come from Galache et al 2008. From the data shown in this figure it is possible to determine a straight line fit using a $\chi^{2}$ routine and establish an average $\dot{P}$ = -1.28s/year over $\sim$11 years. Using Figure 5 of Coe, McBride \& Corbet (2010) this value implies a typical outburst $\dot{P}$ of $\sim$5 times that value.

The X-ray luminosity was calculated from the observed peak pulse amplitude in counts/pcu/s that occurred during the current outburst. This amplitude is converted to luminosity assuming that 1 RXTE count/pcu/s = 0.4$\times 10^{37}$ erg/s at the distance of 60kpc to the SMC (though the depth of the sources within the SMC is unknown and \emph{may} affect this distance by up to $\pm$10kpc - Laney \& Stobie, 1986). The X-ray spectrum was assumed to be a power law with a photon index = 1.5 and an $N_{H}=6\times10^{20}cm^{-2}$. Furthermore it was assumed that there was an average pulse fraction of 33\% for the system and hence the correct total flux is 3 times the pulsed component.
From this we find that the approximate luminosity from our RXTE data in the 3--10keV range is $\sim2 \times 10^{37}$erg/s.
Using the harder IBIS data (15-35 keV) a similar luminosity of $(\sim1) \times 10^{37}$erg/s is found.

 Locating these values for $\dot{P}$ and $L_{x}$ during the outburst on the plot of $log\dot{P}$ versus $logPL^{3/7}$ for SMC sources (Figure 6 of Coe, McBride \& Corbet, 2010) puts SXP756 close to the end of the distribution in the region occupied by two other long period pulsars, SXP892 and SXP1323.

\subsection{IGR J00515$-$7328}

\begin{figure}
\includegraphics[width=60mm,angle=-90]{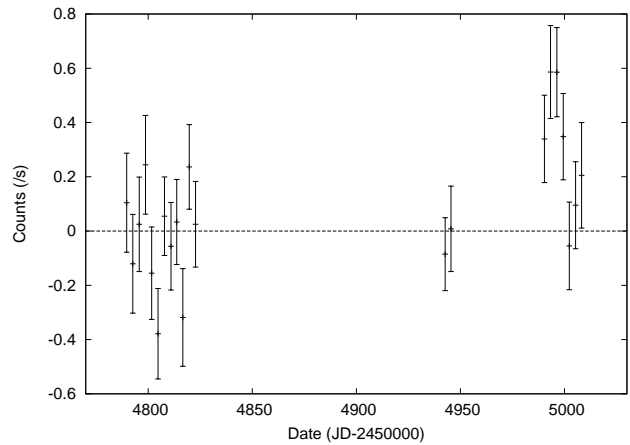}
\caption{INTEGRAL/IBIS 16--35 keV light curve of IGR J00515-7328 }
\label{fig:00515lc}
\end{figure}

A new source,  IGR J00515-7328, was active when the third set of SMC observations started in INTEGRAL revolution 812 (MJD54989). The source continued to be marginally detectable for 3 subsequent revolutions, peaking in flux around MJD54994 (June 12th 2009),
see Figure~\ref{fig:00515lc}

The combined significance across these 4 revolutions is about 5.9 sigma, a robust detection of a faint source. The IBIS source position is RA 00:51:28, Dec -73:30:33 with an error circle of radius 4.2 arcmin. The source had faded below the IBIS detection limit by MJD55000.

\begin{figure}
\includegraphics[width=80mm,angle=-0]{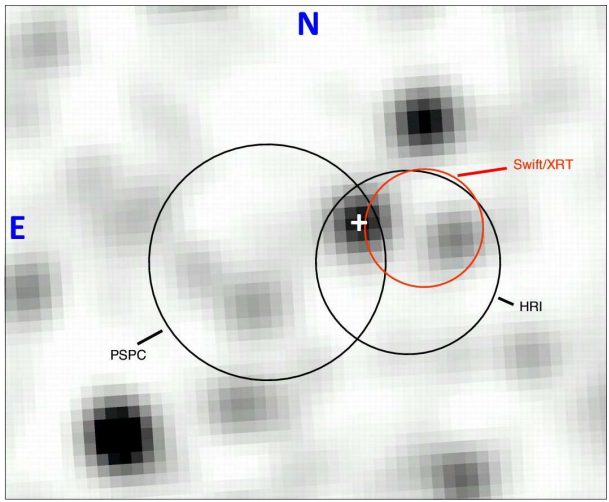}

\caption{A Digitised Sky Survey red-band image showing all the ROSAT and Swift sources known within the IBIS error circle for IGR J00515-7328. The position of 2MASS J00520059-7329255 is indicated by a white cross. The Swift/XRT error circle has a radius of 5.2".}
\label{fig:00515fc}
\end{figure}

A short 500s Swift/XRT TOO was performed on MJD55003, shortly after the source had faded below the detection threshold in IBIS. One faint source was seen within the IBIS error circle at RA=00:51:59, Dec=-73:29:25 with a 5.2" radius error circle. With detection based on ~40 photons, further diagnostic spectral and/or timing analysis was infeasible. A basic extrapolation of the IBIS and XRT fluxes, assuming a typical SMC Be-X spectrum (power law with a photon index of -1.0) and a steady decline of the flux from the source after the IBIS detection, shows the two fluxes to be compatible, but we cannot unambiguously associate the IBIS and XRT detections. The XRT source is very likely associated with a ROSAT PSPC source (Haberl et al., 2000) and a ROSAT HRI source at RA=00:51:59.6 Dec=-73:29:29 (Sasaki et al., 2000), see Figure~\ref{fig:00515fc}.  The flux measured by HRI ($\sim5 \times 10^{34}$ erg/s) is around 20 times lower than the XRT flux ($10^{36}$ erg/s), implying that while the XRT detection was during a decaying outburst, the HRI detection probably represents the quiescent flux from the source.

It is also worth noting that there are 6 XMM sources in the IBIS error circle taken from the XMM Serendipitous Source Catalog (2XMMi version) (Watson et al, 2009). However, none lie close to the area shown in Figure~\ref{fig:00515fc}. Furthermore these XMM sources are restricted to a band across the IBIS error circle suggesting incomplete coverage of this region by XMM.

\begin{figure}
\includegraphics[width=80mm,angle=-0]{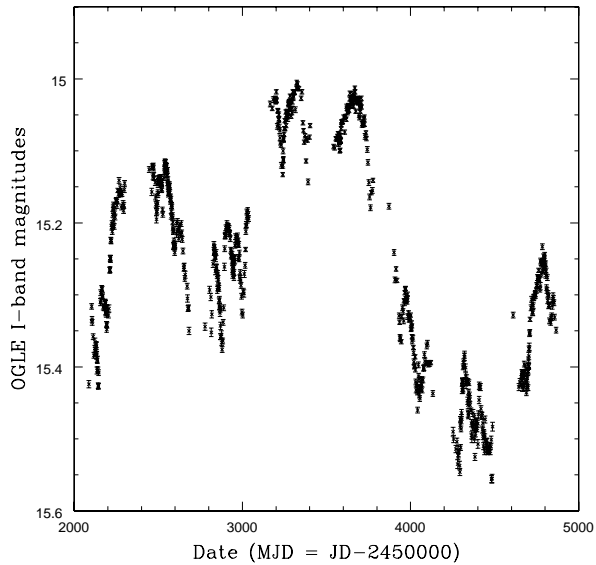}
\caption{OGLE III I-band lightcurve of the proposed optical counterpart IGR J00515-7328 over $\sim$8 years. }
\label{fig:00515ogle}
\end{figure}

2MASS J00520059-7329255 lies 5.6'' away from the XRT position, marginally outside the 90\% error circle and consistent with the ROSAT HRI \& PSPC positions.
This source is also known as [M2002] SMC 21983 (Massey 2002) and has B and V magnitudes of 15.18, typical of Be systems in the SMC. The source to the NW of 2MASS J00520059-7329255 is also in the Mssey catalogue (no: 21891) and has V=16.47 \& B-V=0.42 and is, again, a possible counterpart to a Be X-ray binary system. However, its location well clear of all three error circles makes its association with the ROSAT and Swift objects much less likely. The OGLE light curve (see Figure~\ref{fig:00515ogle}) of 2MASS J00520059-7329255 (OGLE III name SMC 103.4 33693) shows considerable variability on several timescales. Despite this, timing analysis of the de-trended light curve yielded no significant periodicities.

A search of XTE monitoring observations performed on MJD54988 (just before the IBIS detection) and MJD54994 (during the outburst) found only a hint of a 8.29s pulse period in the MJD54994 data. Again, there is no robust way to associate this periodicity with IGR J00515-7328.

Overall, the detection of an X-ray source showing a high level of variability in flux between two epochs, and association with an optical counterpart which is typical of a Be star, suggests that IGR J00515-7328 is another high-mass X-ray binary system within the SMC.

\subsection{SXP6.85}

SXP6.85 (= XTE J0103-728) underwent
a large Type II outburst beginning on 2008 August 10 and the source was consistently
seen in RXTE observations for the following 20 weeks (MJD54688 - 54830) and by INTEGRAL in the period MJD54790 - 54820. Townsend et al (2010a)
presented X-ray timing and spectroscopic
analysis from the Rossi X-ray Timing Explorer (RXTE)
and the INTEGRAL observatory. A
comparison with the Optical Gravitational Lensing Experiment (OGLE) III light curve of the
Be counterpart showed the X-ray outbursts from this source to coincide with times of optical
maximum.

\subsection{SXP11.5 = IGR J01054-7253}

A large outburst from a previously unknown source was seen by INTEGRAL starting MJD54989, lasting $\sim$20d and with a peak 3--10 keV luminosity of $L_{x} = 7\times10^{37}erg/s$. A rapid follow-up observation using the Swift observatory refined the X-ray position and permitted the identification of an associated optical counterpart. Townsend et al (2010b) present X-ray and optical data on this proposed new Be/X-ray binary pulsar, IGR J01054-7253 = SXP11.5.  The optical counterpart was shown to be [M2002] SMC 59977 and OGLE III data on that source presented.  Following the discovery, RXTE began monitoring SXP11.5 approximately 3 time per week for the duration of the outburst (MJD = 55001 - 55063).These data revealed a clear pulse period of 11.48s (Corbet et al., 2009). In addition, a strong Doppler modulation of the 11.48s pulse period was seen and interpreted as an accretion driven spin-up. Model fitting to these data allowed Townsend et al (2010b) to distinguish the spin-up rate from the orbital modulation and identify an orbital period of 36.3$\pm$0.4 days and an eccentricity of 0.28$\pm$0.03 (Townsend et al., 2009).

\subsection{Tentative new sources}

\begin{figure}
\includegraphics[width=70mm,angle=-90]{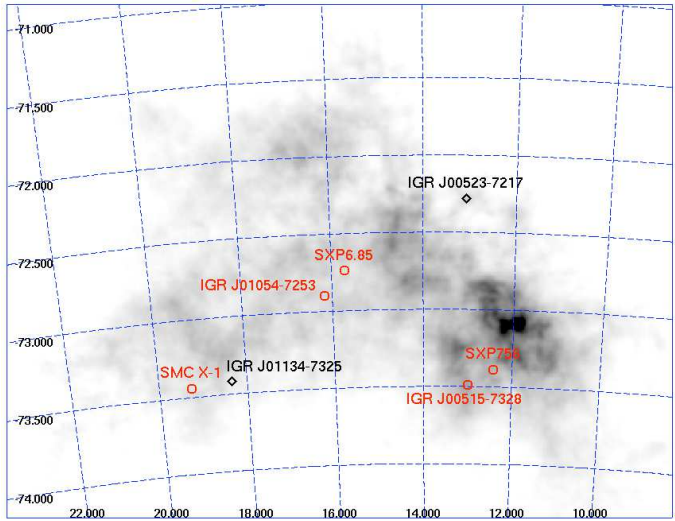}
\caption{Sources detected by INTEGRAL overplotted on an SMC H\,{\small I} column density map (Stanimirovi\'{c}, Staveley-Smith \& Jones 2004). The sources in red are definite source detections discussed in this paper. The two sources in black are the candidate sources listed in Table~\ref{Tab:candidates}. }
\label{fig:tent}
\end{figure}

IBIS 15-35\,keV and JEM-X 3-10\,keV maps on time-scales varying between single revolutions up to five consecutive revolutions were searched for excesses which could indicate potential new sources in the SMC.  A list of candidates is supplied in Table~\ref{Tab:candidates} and mapped on to the SMC in Figure~\ref{fig:tent}.  With the large error circles on these objects, it is not yet possible
to constrain the nature of these sources.  However, their transient nature makes it likely that they are binary systems in the SMC.

\begin{table*}
\caption{Candidate sources in the SMC. Coordinates are in J2000.0. For IBIS detections the fluxes are given in the 15-35\,keV range, while JEM-X fluxes are quoted in the 3-10\,keV range.}
\begin{tabular}{l l l c c c c l}
\hline
Name & RA  & Dec  & Significance &Error radius& IBIS & JEM-X & Flux \\
 & \emph{h\,m\,s} & \emph{d\,m\,s} && arcmin (90\%) & MJD & MJD & (cts/s)\\
\hline
IGR J00523$-$7217 & 00 52 19 & -72 17 06 & 5.3$\sigma$ & 4.7 & 54989-55001& & 0.39$\pm$0.07 \\ 
IGR J01134$-$7325 & 01 13 06 & -73 25 17 & 3.3$\sigma$ & 1.8 & & 54812-54822 &$(1\pm0.3)\times10^{-2}$ \\ 
\hline
\end{tabular}
\label{Tab:candidates}
\end{table*}

One of these candidate sources (IGR~J00523$-$7217) was detected by combining data over a $\sim10$ day period in June 2009 and is spatially consistent with SXP327 \& SXP4.78, known Be/X-ray binaries in the SMC. 
 The region of the SMC in which this potential new source was seen was monitored weekly by RXTE, but pulsations were not detected from either of those sources.


\section{Discussion}

The SMC is turning out to be an exciting nest of X-ray binary pulsars. It is possible to estimate the
number of systems one would expect based upon the relative masses of our galaxy and the SMC. This
mass ratio is approx 50, so with 64 known or suspected systems in our galaxy we would only expect 1
or 2 systems in the SMC. However, Maeder, Grebel \& Mermilliod (1999) have shown that the fraction
of Be stars to B stars is 0.39 in the SMC compared with 0.16 in our galaxy. So this raises the expected number of Be/X-ray systems to approx 3 for the SMC - but we now know of more than »50 such systems in SMC!

Many of these detections have come from XTE, Chandra \& XMM X-ray observations over the last couple
of years (Galache et al. 2008, McGowan et al. 2008, Haberl, Eger \& Pietsch 2008). This large number suggests a dramatic phase of star birth in the past, probably associated with the most recent closest approach between the SMC and the LMC some 0.2 Gyrs ago (Gardiner \& Noguchi 1996). Even more extreme, the very recent work of Naze et al (2003) in just one 20 x 20 arcmin Chandra field identified more than 20 probable Be/X-ray binary systems. Multiplying these numbers up by the »2 x 2 degree size of the SMC, and allowing for $\sim$10\%  X-ray duty cycles, suggests the final number of Be/X-ray binaries could be well in excess of 1,000! These observations of the SMC are not only providing a great sample of HMXBs for study, but are also providing direct insights into the history of our neighbouring galaxy. Since the positional accuracy of Chandra \& XMM is a few arcsec, or less, optical counterpart have been located for many of these potential XRB systems.

The reason for the large number of HMXBs in the SMC probably lies in the history of the Magellanic Clouds. Detailed H1 mapping by Stavely-Smith et al (1997) and Putman et al (1998) has shown a strong bridge of material between the Magellanic Clouds and between them and our own galaxy. Furthermore, Stavely-Smith et al have demonstrated the existence of a large number of supernova remnants of a similar age ($\sim$5 Myr), strongly suggesting enhanced starbirth has taken place as a result of tidal interactions between these component systems. It seems very likely that the previous closest approach of the SMC to the LMC $\sim$100 Myrs ago may have triggered the birth of many new massive stars which have given rise to the current population of HMXBs. In fact, other authors (eg Popov et al 1998) claim that the presence of large numbers of HMXBs may be the best indication of starburst activity in a system.

If the explanation of tidal interactions producing lots of starbirth is correct, it is very likely that there are many more systems waiting to be discovered in the SMC than the 60-100 known at this time. In total, INTEGRAL detected seven sources in the SMC - four of which were previously unknown systems. Combining this with the population of the five new sources detected in the Magellanic Bridge makes this a very productive survey of the SMC region.

Comments on specific SMC sources are:

\begin{itemize}

\item SMC X-1 was overwhelmingly the strongest and most persistent source seen in this survey. It remains the only known supergiant system in the SMC and the results presented here on the changing orbital profile strongly support the earlier work of Trowbridge, Nowak \& Wilms (2007) at lower X-ray energies. It is clear that the superorbital period plays a major role in the obscuration of the emission from this system and the higher energy results presented here should provide further useful constraints on modeling this phenomenon.

\item SXP6.85 is typical of the transient sources detected in this survey. Townsend et al (2010a) attribute this flux variability to the circumstellar disk increasing in size, causing mass accretion
onto the neutron star. Ground based IR photometry and H$\alpha$ spectroscopy obtained during
the outburst were used as a measure of the size of the circumstellar disk and lent support to this
picture. In addition, the folded X-ray light curves seem to indicate complex changes in the geometry
of the accretion regions on the surface of the neutron star, indicative of an
inhomogeneous density distribution in the circumstellar material causing a variable accretion
rate onto the neutron star.

\item The behaviour of SXP11.5 was similar to that of SXP6.85, with the addition of a clear Doppler period shift - something that is often obscured by much larger accretion torques during outbursts. The orbital and spin periods place this system nicely onto the Be/X-ray binary region of the Corbet diagram of binary pulsars. $\dot{P}$ is also consistent with that previously observed in other Be/X-ray binary pulsars in the Magellanic Clouds.

\end{itemize}

The large number of sources seen by INTEGRAL is strongly indicative of the substantial advantage such a wide field instrument has when it comes to determining the extent of a population of transients systems. Though parts of the SMC have been regularly monitored by RXTE over a decade (Galache et al., 2008), the smaller field of view, combined with the collimator response makes it difficult exactly to quantify the duty cycles of these HMXB systems. Other wide field instruments such as the Swift/BAT simply to do not have the combined spatial and flux sensitivity of INTEGRAL/IBIS. Thus the conclusion of the INTEGRAL results presented here is that the total population of HMXBs in the SMC is probably much greater than the 60-70 systems so far catalogued, with consequent implications for the star formation rate in this galaxy. Plus, there is probably a significant population along the Bridge - that may decline in number (along with the age of the stellar population) as one moves from the SMC to the LMC. Future INTEGRAL observations should continue to enhance our understanding of all parts of the Magellanic system.

\section{Acknowledgements}

The OGLE project was partially supported by the Polish MNiSW grant N20303032/4275. LJT is supported by a Mayflower Scholarship from the University of Southampton.

\bsp

\label{lastpage}

\end{document}